# New Perspectives on Classical Electromagnetism

Paul J. Cote

**Abstract**

At present, gauge choice is considered central to electromagnetism. We show here that the gauge topic is generally presented in a manner that often leads to errors and fallacies that impede a deeper understanding of the basic physics. A clearer and more valid formulation of this topic is developed using an approach that preserves both the precise meaning of variables and the distinction among variables. Finally, we show that the issue of gauge choice and its associated problems can be eliminated by recognition of existing physical constraints.

## I. Introduction

In a critique of conventional physics education, Tony Rothman [1] concludes that despite the impressive contributions of physics in the modern world, physics fundamentals are often presented in a jury-rigged and intellectually dishonest fashion so that the entire enterprise now resembles Bruegel's chaotic Tower of Babel. Rothman sees the origin of the problem as the perceived need for shortcuts to simple instructional results. One of the consequences of these simplifications is a loss of basic understanding of the underlying physics.

Our experience with the gauge concept in classical electromagnetism suggests that groupthink is another factor in the general problem described by Rothman. Groupthink is characterized by an intolerance to criticism of the orthodox view, which acts to suppress self-correction and perpetuate the chaos. An indication of the uncritical acceptance of the current gauge formulation is that the familiar Lorenz gauge, $\nabla \cdot A = -(\partial \varphi_C / \partial t)/c^2$, has universally been attributed to H.A. Lorentz for generations. It is only recently that the author of the concept is recognized as L. Lorenz [2]. This correction can be found in the latest edition of Jackson's textbook [3], for example.

At present, gauge choice is the centerpiece of electromagnetism. Electromagnetism is considered a paradigm for gauge theories where the freedom to choose arbitrary functions for a gauge is seen as a convenience in problem solution. While it can be viewed as a convenience in a limited context, the careless application of this term has contributed to fallacies and confusion. The gauge approach has evolved into a number of ad-hoc rules that give the right answer: certain types of problem require certain gauge choices. In the authors' experience, if one presses the matter as to basic reasons for a given gauge choice, the answer is usually a matter of diverse opinions as to which of the many published papers on the subject has the best answer. In our view, this diversity of opinions is another indication of a problem.

The issue of gauge choice arises in the discussion of the vector potential. The conventional view is that only the curl of the vector potential has meaning through its connection to the magnetic field, so the vector potential itself has little physical significance and always requires an arbitrary gauge choice for its full definition. Konopinski [4] challenges that view. He demonstrates that the vector potential has physical meaning in terms of a field momentum. His definition is analogous to the definition of the electric field in terms of the force on a test charge q: The field momentum is defined and measured by the momentum, $qA/c$, imparted to the charge as the source for $A$ is slowly applied. This gives $A$ measurability, at least in principle for simple cases. Konopinski's demonstration of the physical reality of the vector potential is important for two reasons. First, it satisfies the normal requirement that physics deals with quantities and relationships that have physical meaning and are testable. The other reason is that once the physical reality of the vector potential is recognized, the notion that one can always add an arbitrary scalar component to the vector potential must be qualified in view of physical restrictions imposed by this reality. An example is that real dynamic field variables must propagate as an electromagnetic wave. Another physical constraint is that the vector potential is also defined by its relationship to the induced field.

Konopinski is not alone in his view of the physical reality of the vector potential. Feynman el al.[5] discuss the results of the Bohm-Aharanov experiment that illustrate the physical reality of the vector potential, in the context of quantum mechanics. Like Konopinski, Feynman et al. disagree with the current textbook presentations on the topic of the vector potential; they specifically complain about the prejudice that exists in the physics community on this subject. In that sense, the present study is an extension of their analyses.

A common misconception is that only the vector potential is undefined. Actually, in the present formulation, in the presence of dynamic Coulomb fields, neither the Coulomb field, nor the induced electric field is defined. This basic premise of the gauge based approach conflicts with the normal requirement that physics deal with defined, physically meaningful quantities.

A resolution of these issues is developed in the following. The major part of this report examines the consequences of incorporating existing physical requirements on the treatment of gauge choices.

## II. Demonstration of gauge fallacies

A contributing source for the gauge-related fallacies is the failure to distinguish among variables. In the standard formalism, the labels $E$, $A$, and $\varphi$ are each used to represent a variety of different variables. We apply subscripts throughout the following to clearly preserve the meanings of the variables.

We begin with some elementary notes. According to the Helmholtz theorem, any physically meaningful vector can be written as a sum of a gradient of a scalar and a curl of a vector. A gauge choice is required in cases where one needs to obtain quantitative expressions for variables that are incompletely defined. One chooses a gauge that produces the simplest, least cumbersome form for relationships among variables, in analogy with the choice of a zero for the Coulomb potential. Alternative choices of gauge cannot affect the fundamental physics.

A. **Hidden gauge**

The magnetic field $B$ is given by,

$$B = \nabla \times A. \tag{1}$$

According to the Helmholtz theorem, the general expression for the vector potential, $A$, is then given by

$$A = \nabla \times F_A + \nabla \varphi_A. \tag{2}$$

Equation (1) defines $A$ to within an arbitrary function, $\nabla \varphi_A$, so that a gauge choice is required if no other information is available. A non-zero divergence of a vector implies the existence of a scalar field associated with that vector. The following summarizes the usual development of the gauge based approach and illustrates how it is a source of confusion.

Faraday's law is originally expressed as a line integral of the induced field, $E_I$, around a closed path, which leads to the relationship, $\nabla \times E_I = -\partial B / \partial t$. Since the Coulomb field is always derivable from a scalar potential, $E_C = -\nabla \varphi_C$. When these two electric fields are present together, the total field is $E = E_I + E_C$, giving $\nabla \times E = -\partial B / \partial t$. Applying Eq.(1) gives $\nabla \times (E + \partial A / \partial t) = 0$. Thus, $E + \partial A / \partial t = -\nabla \varphi$, so that $E$ is given by

$$E = -\nabla \varphi - \partial A / \partial t. \tag{3}$$

This approach presents two problems. First it artificially couples the two basic electric fields as if the principle of superposition does not apply. This is a coupling of two entirely different electric fields. Second, it leaves the impression that $A$ is the only variable requiring a gauge choice. Regarding the second item, if one preserves the distinction among variables, it is seen that there are actually two gauges to consider. A clearer development of the basic equations is offered next.

Adhering to the original form of the Faraday law, which relates only to the induced field, $E_I$, applying Eq.(1) gives

$$\nabla \times E_I = -\partial B / \partial t = -\nabla \times \partial A / \partial t. \tag{4}$$

As with Eqs. (1) and (2), the curl provides an incomplete definition of $E_I$ since

$$\nabla \times E_I = \nabla \times (E_I + \nabla \varphi_I). \tag{5}$$

So, the general expression for $E_I$ is,

$$E_I = -\partial A / \partial t - \nabla \varphi_I. \tag{6}$$

Therefore, the general expression for total electric field is,

$$E = E_C + E_I = -\nabla(\varphi_C + \varphi_I) - \partial A / \partial t, \tag{7}$$

with $A$ given in general form by Eq.(2). So, comparing Eqs.(3) and (7) shows that, in the general case, $\varphi$ is not the Coulomb potential as conventionally assumed, but is the sum of two scalar fields, $\varphi = \varphi_C + \varphi_I$. So the common textbook assumption that $\varphi$ is just the Coulomb potential represents an adoption of a hidden gauge choice, $\varphi_I = 0$ which requires $E_I = -\partial A / \partial t$. Similarly, Eq.(6) shows that the standard practice of employing

$$E_I = -\partial A_S / \partial t, \qquad (8)$$

represents the same hidden gauge choice ($\varphi_I = 0$). The subscript on $A_S$ denotes the adoption of the standard gauge. (Note that $\varphi_I = 0$ does not necessarily mean $\nabla \cdot E_I = 0$ because the vector potential is still described by Eq.(2).) *This hidden gauge choice is always made in electromagnetism in either of the two ways given above, so $\varphi_I = 0$ is always the implicit "standard gauge" in the textbook approach to electromagnetism* [3,4,5,6]. It follows that any treatment of gauge choice that ignores this fact is susceptible to error and should be viewed with caution.

Generally, different gauge choices result in different vector potentials. The "true" Coulomb gauge, $\nabla \cdot A_C = 0$, requires use of Eq.(6) which is not generally employed. So, the only gauge used in the standard approach is the standard gauge. This is the case despite the general acceptance that electromagnetism serves as a gauge theory paradigm. We will review examples of the standard gauge in treatments of the wave equations, the four-vector formulations, and the Lienard-Weichert potentials. Why is the true Coulomb gauge not employed ? We include true Coulomb gauge results in later discussions to help answer this question. The role of the hidden gauge on the formalism is examined in the following section.

**B. Hidden gauge and Gauss' law**

Gauss' law for the basic dynamic $E_C$ and $E_I$ fields is given by

$$\nabla \cdot E = \nabla \cdot (E_C + E_I) = \rho / \varepsilon. \qquad (9)$$

Inserting Eq.(7) into Eq.(9) gives the general expression for Gauss' law,

$$\nabla \cdot E = -\nabla^2 (\varphi_C + \varphi_I) - \partial (\nabla \cdot A) / \partial t = \rho / \varepsilon, \qquad (10)$$

where $A$ is given by Eq.(2). Applying the hidden gauge, $\varphi_I = 0$, gives

$$-\nabla^2 \varphi_C - \partial (\nabla \cdot A_S) / \partial t = \rho / \varepsilon. \qquad (11)$$

Since $\nabla \cdot A_S = \nabla^2 \varphi_A$, Eq.(11) can be rewritten as

$$\nabla^2 (\varphi_C + \partial \varphi_A / \partial t) = \nabla^2 \varphi = -\rho / \varepsilon. \qquad (12)$$

Consider the corresponding true Coulomb gauge ($\nabla \cdot A_C = 0$, $\varphi_A = 0$) case to complete the results for the two simplest gauge choices ($\varphi_I = 0 \, or \, \varphi_A = 0$). From Eq.(10),

$$\nabla^2(\varphi_C + \varphi_I) = \nabla^2 \varphi = -\rho/\varepsilon. \tag{13}$$

Note the similarity of Gauss' law in the two gauges, Eqs.(12) and (13). Both express the same physics contained in the hidden law, but in different forms. Both say that the sum of two dynamic scalar fields obeys Gauss' law. Consequently, the same electrostatic Laplacian expression applies to sum of the dynamic potentials in both gauges.

There is a simple relationship between these two gauge scalar potentials. This relationship is obtained by equating the above Gauss' law expression in the two respective gauges, the standard gauge, $\varphi_I = 0$ and the true Coulomb gauge, $\varphi_A = 0$. The result is,

$$\varphi_I \leftrightarrow \partial \varphi_A / \partial t. \tag{14}$$

Equation (14) reflects the fact that both $\varphi_I$ and $\partial \varphi_A / \partial t$ represent the same real, induced scalar potential in the context of the two gauge choices.

### C. Hidden gauge and the peculiarity of the Coulomb gauge for dynamic fields

Jackson's textbook [3] offers a demonstration that the standard treatment of gauge choice leads to the requirement that dynamic Coulomb fields propagate instantaneously. This demonstration is Jackson's attempt to warn the physics community of a fundamental problem with the treatment of gauge choices, which he labels a "peculiarity".

We reproduce Jackson's demonstration here: First, ignore the distinction between $A_C$ and $A_S$. Next, apply $E_I = -\partial A / \partial t$. Finally, invoke the Coulomb gauge, $\nabla \cdot A = 0$, in the dynamic form of Gauss' law, Eq.(9). This gives,

$$\nabla^2 \varphi_C = -\rho/\varepsilon. \tag{15}$$

Given that $\varphi_C$ is a basic field, and cannot be considered a sum of retarded fields in the manner described in section B, Eq.(15) indeed requires that dynamic Coulomb fields propagate instantaneously, which is a physical impossibility. What is wrong with this proof? Jackson's textbook suggestion is that the problem originates from the basic unreliability of electromagnetic potential formulations. It is shown in the following section that the problem originates, instead, from the inappropriate treatment of gauge choices.

**D. Overlooked law of physics**

The usual derivation of the dynamic form of Gauss' law, Eq.(9), involves nothing more than inserting the sum of the dynamic fields into the static expression for the divergence of the Coulomb field. That is not a legitimate procedure, because, a priori, there is no reason to expect that the static equation for a Coulomb field should also apply to the sum of dynamic fields. The fact that the result has proven, in practice, to be a valid equation is beside the point. (This is as a good example of Rothman's observation regarding the presentation of oversimplified results that neglect the underlying physics.)

Given its proven validity, Eq. (9) must contain an overlooked physical law. Under dynamic conditions, the Coulomb field component, $E_C$, of the total electric field, $E$, cannot be treated as if it propagates instantaneously because of retardation effects, so $E_C$ cannot obey Gauss' law. The overlooked law can be expressed as: *A dynamic Coulomb field always induces a self-correcting dynamic scalar component for the induced field, $E_I$, so that the total electric field obeys Gauss' law.* Consequently, $E$, which is the sum of retarded Coulomb and induced fields, propagates instantaneously from central sources, as required by Eq.(9). In other words, $E$ must behave as if it were propagating as a quasi-static field with all field lines originating and terminating on the instantaneous position of moving charges. Thus, the dynamic form of Gauss' law requires that the sum of the two basic retarded fields is a non-retarded field.

This result is already familiar to us from the elementary case of the Lienard-Weichert retarded potentials for a moving point charge. For a moving charge, neither $E_C$ nor $E_I$ at a distant point in space can originate from the present location of the moving charge because of retardation effects arising from the finite speed of light. The total field, $E$, while no longer spherically symmetric, is radial and always tracks the instantaneous position of the moving point charge. Thus, $E$, obeys Gauss' law. This effect in the Lienard-Weichert potentials is more than just an isolated curiosity. It is arguably the most fascinating phenomenon in electromagnetism. The principal reason that the central role of this "curiosity" in Gauss' law, (Eq.9), has been ignored to date is that the present gauge approach masks its presence.

As a reminder, the need for Eq. (9) and the overlooked law rests on charge conservation. This can be seen by taking the time derivative of Gauss' law, Eq.(9), which gives,

$$\nabla \bullet J_T + \partial \rho / \partial t = 0. \tag{16}$$

where $J_T$ is the true current. The solenoidal total current is given by $J_{TOT} = J_T + \varepsilon \, \partial E / \partial t$. So the real basis for the dynamic form of Gauss' law is the requirement of charge conservation.

As an elementary example of the importance of the overlooked law, it explains the remarkable fact in A.C. circuits that the current instantaneously tracks the source of the current all along the conductors, regardless of dimension, and regardless of how rapidly the source may vary. A question that may arise in an elementary physics class is: How does nature accomplish this without violating relativity? The explanation requires recognition of the overlooked law. The displacement current fields and the fields that drive the charges in a conductor are both obtained from sums of retarded $E_C$ and induced $E_I$ fields, and these sums always propagate instantaneously, so that current continuity is preserved.

Returning to the overlooked law, the situation for $E_I$ is similar to that of the vector potential. Faraday's law offers an incomplete description of the induced field because it only defines $E_I$ via a closed line integral. Consequently, any scalar component is left undefined. A complete definition of $E_I$ is available in terms of a precise mathematical expression of the overlooked law that is, in effect, lying in plain sight: Recognizing the scalar wave equation for the retarded Coulomb field as a fundamental mathematical necessity [5], as well as a physical requirement for any real retarded electromagnetic field, we invoke the familiar wave equation for the Coulomb potential,

$$\nabla^2 \varphi_C - (\partial^2 \varphi_C / \partial t^2)/c^2 = -\rho/\varepsilon. \tag{17}$$

Comparing this equation with the general expression for Gauss' law, Eq.(9) gives

$$\nabla \cdot E_I = (\partial^2 \varphi_C / \partial t^2)/c^2. \tag{18}$$

Equation(18) connects the mathematical source term for a retarded field in the differential equation, Eq.(17), with a physical source for the induced field, $E_I$. As a consistency check, note that in the absence of dynamic Coulomb fields, both Eq. (18) and the general Gauss' law, Eq.(9), reduce to the same result, $\nabla \cdot E_I = 0$.

The source term in Eq. (18) is a continuous distribution of central sources. An elementary example of this dynamic effect is given in Chapter 21 of Feynman et al [5] where retardation transforms a Coulomb point source into an continuous distribution of sources.

With Eq.(18) accepted as a fundamental equation, one now has a complete characterization of $E_I$ as well as $E_C$ (via Eq.(9)). Basic physics must require that the physically real Coulomb fields propagate at the speed of light, regardless of gauge choice. Combining Eq.(18) with the standard gauge expression, $E_I = -\partial A_S / \partial t$, gives $\partial/\partial t \{ \nabla \cdot A_S + (\partial \varphi_C / \partial t)/c^2 \} = 0$, so that,

$$\nabla \cdot A_S + (\partial \varphi_C / \partial t)/c^2 = 0. \tag{19}$$

(The possibility of a static term in Eq.(19) is unphysical since this is strictly a dynamic effect. We will see in section F that this solution also satisfies the requirement for the wave equation for the vector potential.) Eq.(19) is the same as the familiar Lorenz condition, but now it not a condition. Instead, it is obtained from laws of physics. Note that Eq.(19) is the same as the overlooked law (Eq.(18)) but expressed in the standard gauge. In both of these expressions, the dynamic Coulomb field provides the common, physically real distribution of central sources for the fields.

These results illustrate the two points. First, given an initial gauge choice, the second gauge choice is restricted by the laws of physics. Second, the Lorenz condition is more than just a condition. It follows from physical requirements. For clarity, we will refer to the derived equation, Eq.(19), as the "Lorenz equation".

The conclusion regarding the role of retarded fields and current continuity in establishing the physical necessity of Eqs.(17), (18), and (19) follows from the argument that the basic Coulomb fields must propagate as a wave travelling at the speed of light. We can arrive at the same conclusion in another way using the following pair (Eqs (20) and (21)) of general retarded field equations [3,5,6] for the vector and Coulomb potentials.

$$A_S(1,t) = \int \frac{\mu J_T(2,t_r)}{4\pi r_{12}} dV_2, \qquad (20)$$

and,

$$\varphi_C(1,t) = \int \frac{\rho(2,t_r)}{4\pi\varepsilon r_{12}} dV_2, \qquad (21)$$

where the fields are evaluated at position 1 and present time t, from sources located at position 2 and retarded time $t_r = t - r_{12}/c$.

Equations (20) and (21) are generally presented [3,5,6] as solutions to the wave equations that are derived from Maxwell's equations with the incorporation of the arbitrary Lorenz condition. Equation (21), however, can be viewed instead as a modification of the corresponding static expressions using the requirement that electromagnetic fields must propagate at the speed of light in free space. It is therefore a first principles result that is independent of gauge. It can verified by direct differentiation that Eq.(21) satisfies the wave equation for $\varphi_C(1,t)$; the calculation is straightforward, but a bit tedious, and involves differentiation of $\varphi_C(1,t)$, the chain rule, and the relationship, $\nabla_1^2(1/r_{12}) = -4\pi\delta^3(r_{12})$. It follows that Eq.(18) must also apply in order to satisfy Gauss' law, so the missing law of physics is gauge invariant.

The vector potential, on the other hand, is dependent on gauge choice. The vector potential, $A_S$, is completely defined in the standard gauge by Eq. (20).

It is easy to see that Eq.(20) also applies in the true Coulomb gauge case, provided $J_{TOT}$ is used in place of $J_T$. We use this modified form in the discussion of the true Coulomb gauge in part IV.

Given that all relevant variables are now completely defined in the standard gauge by Eqs. (20) and the gauge invariant Eq.(21), and that their associated fields all satisfy Maxwell's equations, as well as the overlooked equation, Eq.(18), it is proper to recognize this pair as the core equations of electromagnetism. This recognition is consistent with the conclusions expressed by Konopinski[4] and Feynman et al[5] that the potentials should form the basis of classical electromagnetism, just as they do in quantum mechanics.

Returning to Jackson's discussion of the Coulomb gauge peculiarity, when he makes the implicit initial gauge choice, $\varphi_I = 0$, by employing the equation, $E_I = -\partial A_S/\partial t$, physics then requires that $\nabla \cdot A_S = -(\partial \varphi_C/\partial t)/c^2$. Any other assignment for $\nabla \cdot A_S$ is a violation of the laws of physics. That is the reason for Jackson's unphysical result. Jackson's demonstration breaks one of the ad-hoc rules: In the standard treatment, one cannot employ the Coulomb gage for any problem where dynamic Coulomb fields exist because it gives the wrong answer.

More generally, given that the implicit gauge ($\varphi_I = 0$) is always invoked as described above, gauge choice is, in practice, never a consideration in the standard treatment of electromagnetism. This point could have been consigned to a footnote in textbooks, with no further discussion of the matter. The central role of gauge choice and its associated problems would thus have been eliminated. It will be shown in part VI, even this free choice in the standard gauge is removed if one recognizes the constraints imposed by Coulomb's law.

We discuss the implications of these results in the remainder of this report to provide further insights. We include results for the true Coulomb gauge to illustrate some of the above points and explain why this gauge is not used.

### E. In the absence of dynamic Coulomb fields

Consider cases where dynamic Coulomb fields are absent. There are no retarded Coulomb fields. The standard gauge is selected in practice, since $E_I = -\partial A_S / \partial t$ is assumed, so Eq.(19) applies. (The ad-hoc rule is that the Coulomb gauge must be invoked here.) So in the absence dynamic Coulomb fields, the "Coulomb gauge" choice, $\nabla \cdot A_S = 0$ is not a gauge choice. Instead, it is a physical requirement because $\partial \varphi_C / \partial t = 0$.

The true Coulomb gauge $\nabla \cdot A_C = 0$ gives the same result. So, both vector potentials now have identical curls and divergences, which means $A_S = A_C$, and, given $E_I = -\partial A_S / \partial t = -\partial A_C / \partial t - \nabla \varphi_I$, it follows that $\nabla \varphi_I = 0$. All fields are fully defined and are identical in both gauges. Thus, there is no gauge choice in the absence of dynamic Coulomb fields. Thus, the term "Coulomb gauge" is always used improperly in these cases since the resulting fields are independent of gauge choice.

As an elementary example, consider a circular metal ring in a steady state, time varying magnetic field (Fig. 1a). This is similar to the betatron configuration. Recall that potential differences are given by a line integral of $E_C$, while emf's are given by a line integral of $E_I$. (This is also an illustration that these basic fields need to be treated individually in the solution of real problems.) Given the ring symmetry and field uniformity, there can be no potential differences, so the current changes are actually generated by the uniform distribution of an emf along the ring. Thus, there are no induced scalar potentials, and no free choice for $\nabla \cdot A$. A physically real vector potential must be solenoidal in this case.

Another elementary example is electromagnetic radiation in free space, remote from any Coulomb sources. Again, only solenoidal fields exist. In classical electromagnetism, the radiation fields and their sources propagate through space together at the speed of light by means of successive generations of time varying closed loops of magnetic fields which induce closed loops of displacement currents, which, in turn, induce new magnetic field loops, continuing ad infinitum. Electromagnetic radiation is always addressed in terms of the convenient choice of the radiation gauge, $\nabla \cdot A = 0$, but, again, in reality, there is no gauge choice.

It should be noted there are no dynamic Coulomb fields in Konopinski's [4] definition of the vector potential or in the Bohm-Aharanov experiment [5]. So the discussions of these topics need to be modified to incorporate the fact that there is no gauge choice so that the vector potentials are fully defined.

**F. Wave equations in the two gauge choices**

The wave equations are generally treated in the standard gauge ($\varphi_I = 0$), but we will also discuss the true Coulomb gauge form ($\varphi_A = 0$). The general expression for wave equation for the vector potential for both arbitrary gauge choices is obtained using Ampere's law,

$$\nabla \times B = \nabla \times \nabla \times A = \mu J_{TOT} = \mu(J_T + \varepsilon \partial E / \partial t), \quad (22)$$

where $J_T$ is the true current, and $\varepsilon \partial E / \partial t$ is the quasi static sum of the two retarded displacement currents. Applying Eq.(7) to Eq.(22) gives the general wave equation (prior to any gauge selection):

$$\nabla(\nabla \cdot A) - \nabla^2 A = \mu J_T - \mu\varepsilon \partial \nabla(\varphi_C + \varphi_I) / \partial t - \mu\varepsilon \partial^2 A / \partial t^2. \quad (23)$$

The vector potential wave equation in the standard gauge, $\varphi_I = 0$, is obtained directly from Eq.(23), giving the familiar wave equation for the vector potential, $A_S$,

$$\nabla^2 A_S - \mu\varepsilon \partial^2 A_S / \partial t^2 = -\mu J_T + \nabla(\nabla \cdot A_S + (\partial \varphi_C / \partial t)/c^2). \quad (24)$$

(Again, whenever these expressions are employed, there is as implicit assumption of the standard gauge.)

Applying the solution, Eq.(19), gives the wave equation for the non-solenoidal $A_S$,

$$\nabla^2 A_S - \mu\varepsilon \partial^2 A_S / \partial t^2 = -\mu J_T, \quad (25)$$

This is also a consistency check on Eq.(19), since all real physical fields, including $A_S$, must propagate as electromagnetic waves.

Note that this derivation differs significantly from the usual approach, where the arbitrary Lorenz condition is used to effect a decoupling of the scalar and vector wave equations. Equation(25) is obtained here on the basis of physical requirements; there is no decoupling because the scalar wave equation is gauge invariant.

A key point is that the general solution to Eq.(25) is given by the integral in Eq.(20) over the source current, which, in this case, is the true current, $J_T$. Typically, true currents refer to moving charges in conductors. Figure 1b is a schematic of a circuit that contains a Coulomb field. The portion of the circuit represented by $J_T$ is not a complete circuit, and therefore not solenoidal. The vector potential, $A_S$, generated by that source is obtained by the integral over $J_T$ using the potential equation, Eq.(20). If one calculates the divergence of $A$ in Eq.(20), one finds that solenoidal sources ($\nabla \cdot j = 0$) produce solenoidal fields and non-solenoidal sources ($\nabla \cdot j \neq 0$) produce non-solenoidal fields. So the physical meaning of the vector potential obtained in the standard gauge is that $A_S$ is a component of the total vector potential that would be obtained by integrating over the entire solenoidal current loop. Hence, it cannot be solenoidal.

We now discuss the corresponding wave equations in the true Coulomb gauge formulation. Eq.(8) no longer applies for $E_I$. Instead, the correct expression is obtained from Eq.(6). Assuming the true Coulomb gauge,

$$E_I = -\partial A_C / \partial t - \nabla \varphi_I. \tag{26}$$

The true Coulomb gauge expression for the wave equation is obtained from Eq.(23) using $\nabla \cdot A_C = 0$ from Eq.(2) where $\varphi_A = 0$, which gives,

$$\nabla^2 A_C - \mu\varepsilon \partial^2 A_C / \partial t^2 = -\mu J_T + \mu\varepsilon \partial \nabla(\varphi_C + \varphi_I) / \partial t. \tag{27}$$

The scalar wave equation for $\varphi_C$ obtained from Eq.(21) also applies here because it is independent of gauge choice. Combining Eqs. (13), (17), and (26), gives,

$$\nabla \cdot E_I = -\nabla^2 \varphi_I = \partial^2 \varphi_C / \partial t^2 / c^2. \tag{28}$$

Thus, the overlooked, law of induction, Eq.(18), is obtained in both formulations.

Returning again to Jackson's demonstration that Coulomb fields propagate instantaneously in the Coulomb gauge, note that in the true Coulomb gauge, neither the scalar nor the vector potentials propagate instantaneously.

Equation (28) describes how the dynamic Coulomb field now induces the scalar $\varphi_I$ directly in $E_I$ in the Coulomb gauge formulation, rather than indirectly via the vector potential, as in the standard gauge. And it explains more clearly the meaning of Eq.(14).

As for the physical meaning of the vector potential in the true Coulomb gauge, note that the displacement current arises from the time derivative of the quasi-static, $\nabla\varphi$ in Eq.(27). Thus, $\nabla\varphi$ behaves as if it were an instantaneously propagating longitudinal field, so its time derivative is an instantaneously propagating displacement current. The complete current loop, $J_{TOT} = J_T - \varepsilon \partial \nabla\varphi / \partial t$ serves as the solenoidal current source for the solenoidal vector potential, $A_C$. In this case the solution for $A_C$ obtained (in principle) from Eq.(20) using $J_{TOT}$ as the source current.

The wave equations in the two gauge contexts, Eqs. (25) and (27), express the same physics. The divergence of both returns the continuity equation, and the curl of both returns the wave equation for the magnetic field, $B$. (The reason that the two different vector potentials $A_S$ and $A_C$ give the same $B$ is that there is no net contribution from the scalar component.) Both vector potential wave equations give the same wave equation for the induced field, $E_I$, as they must if $E_I$ is gauge invariant. And, one can complete the set of basic equations with the gradient of Eq. (17), which again gives the original wave equation for the Coulomb field, as it must since it is also gauge invariant.

Finally, it should be noted that the freedom to assign the real, induced scalar field to either $\varphi_A$ or $\varphi_I$ is directly related to the Lorenz gauge transformation where one is free to add a gradient of a scalar to the vector potential as long as one subtracts a compensating term from the scalar potential. So, the Lorenz transformation is not just a mathematical trick, it has physical significance.

### III. Four- vector formulation

The four-vector formulation for electromagnetism is only given in the standard gauge in textbooks. We modify the standard four-vector notation in the following to more precisely reflect this standard gauge choice. We also discuss the four vector formulation in the true Coulomb gauge, as an illustration that the standard gauge is, not unique in providing four-vector expressions. The basic requirement for the four-vector formulation follows from the fact that Maxwell's equations are invariant to a Lorentz transformation.

Following the $c=1$ notation in Feynman et al [5], the four- vector has the form,

$$a_\mu = (a_t, a_x, a_y, a_z). \tag{29}$$

The four-vector divergence is given by

$$\nabla_\mu a_\mu = \partial a_t / \partial t + \nabla \cdot a, \tag{30}$$

and, the four-vector Laplacian is given by

$$\Box^2 = \partial^2 / \partial t^2 - \nabla^2. \tag{31}$$

Preserving the distinction among vector potentials, the standard gauge potential is given by,

$$(A_S)_\mu = (\varphi_C, A_S). \tag{32}$$

Using the current vector,

$$j_\mu = (\rho, (J_T)_x, (J_T)_y, (J_T)_z), \tag{33}$$

gives,

$$\Box^2 (A_S)_\mu = j_\mu / \varepsilon, \tag{34}$$

and,

$$\nabla_\mu j_\mu = 0. \tag{35}$$

Equation(19), the Lorenz equation, is given by,

$$\nabla_\mu (A_S)_\mu = 0. \tag{36}$$

The true Coulomb gauge results can also be expressed in four-vector form. The true Coulomb gauge wave equation developed earlier gives:

$$\Box^2 (A_C)_\mu = j_\mu / \varepsilon, \tag{37}$$

where,

$$(A_C)_\mu = (\varphi, A_C), \tag{38}$$

and,

$$j_\mu = (\rho, (J_T - \varepsilon \partial \nabla \varphi / \partial t)_x, (J_T - \varepsilon \partial \nabla \varphi / \partial t)_y, (J_T - \varepsilon \partial \nabla \varphi / \partial t)_z), \tag{39}$$

where $\varphi$ is the sum of scalar potentials. In this case,

$$\nabla_\mu j_\mu = \partial \rho / \partial t. \tag{40}$$

Thus, we have shown that the true Coulomb gauge formulation also produces the basic equations of electromagnetism, complete with their four-vector expressions. We will return to the question of the special role for the standard gauge in the four-vector formulation and throughout electromagnetism.

### IV. Picture of reality

Many of the points made in Part II are summarized in schematic form in Figure 1. It compares the two general categories of problems encountered in electromagnetism. Figure 1a represents cases where external dynamic Coulomb fields are absent and Figure 1b represents cases where dynamic Coulomb fields are present. In Figure 1a, either the true current, $J_T$, or the displacement current $J_D = \varepsilon \partial E / \partial t$ forms a closed loop so that the resulting vector potential field is solenoidal and the gauge concept does not apply

Figure 1b corresponds to more general circuits where both types of current, $J_T$ and a displacement current $J_D$ exist together to form a closed loop ($A_C \neq A_S$). The overall circuit geometry is similar, as shown in the two figures, so the fields are similar in both. It should be clear from this diagram that the true Coulomb gauge vector potential $A_C$ is unique and solenoidal. As discussed, the $A_S'$ field is a non-solenoidal component of $A_C'$. (A prime is added to the field terms to reflect the actual differences in total current configurations in Figures 1a and 1b.)

A major difference between Figures 1a and 1b is that total current in Figure 1b extends throughout space. This is actually a complicated retarded field generated by the moving charges on the capacitor plates. To understand even these most elementary problems, one needs at least an awareness that retardation effects are at the heart of electromagnetism. The basic feature of retarded fields is that their sums, $E = E_C + E_I$, are always quasi-static fields. These $E$ fields originate and terminate at the instantaneous position of the charges on the capacitor plates, so that the continuity condition for the current applies.

These examples also illustrate the reason why the standard gauge is the preferred gauge. Consider the schematic in Figure 1b. In order to compute $A_C$, one needs to integrate the current over the entire circuit, which includes the quasi-static displacement current that extends throughout space. (The displacement current also requires the solution for the scalar and vector potentials for its calculation.) That is generally an intractable task. Compare that to the standard gauge case where the true current, $J_T$, is usually confined to a highly restricted region in space within a conductor so that calculations for $A_S$ are manageable. (This point is not new since the Lorenz condition was introduced in the first place because it allowed simplified treatment of Eq.(24).)

Consider the example of the Lienard-Weichert vector potential where the standard gauge is assumed. In that case, $J_T$ is a point source of current, so that $A_S$ is readily computed. There is no need to integrate over the remaining highly complicated displacement current that extends through space. Attempting the direct solution for the Lienard-Weichert potentials in the true Coulomb gauge is not feasible.

### V. Coulomb's Law

As discussed above, a good reason for the general use of the standard gauge in the presence of dynamic Coulomb fields is that it is generally the only practical choice for problem solving. A more fundamental reason for restricting the gauge choice to the standard gauge can be found in Coulomb's law. If one requires that the only real, physical field for an isolated point charge in its rest frame is the Coulomb field, $E_C$, then the only gauge that satisfies that requirement is the standard gauge.

Consider the four vector formulation discussed above. It is well-known [5] that a Lorentz transformation of the field for a point charge from its rest frame to a moving frame produces exactly the dynamic Lienard-Weichert potentials that were computed in the standard gauge (i.e., $E_I = -\partial A_S / \partial t$, $\nabla \bullet A_S = -(\partial \varphi_C / \partial t)/c^2$).

Alternatively, suppose the fields could have been obtained in another gauge. Then the Lorentz transformation from the moving frame to the rest frame would necessarily result in an unphysical scalar field, $\varphi \neq \varphi_C$ and a vector potential, $A \neq 0$, in the rest frame so that, from Eq. (7), $E = -\nabla(\varphi) - \partial A/\partial t = E_C$. This violates the requirement that the only real field in the rest frame of a point charge is the Coulomb field.

So, in view of the constraints imposed by Coulomb's law and relativity, the Lorentz equation becomes a requirement of Coulomb's law and special relativity. And it follows that the missing law, Eq.(18) is required as well, since the two are related by a time derivative. This result is consistent with all the conclusions drawn earlier for the standard gauge. The only difference is that now gauge choices are eliminated because the standard gauge is a requirement of Coulomb's law.

## VI. Summary and Conclusions

Examination of the standard textbook treatment of gauge choice in electromagnetism shows how shortcuts that gloss over the basic physics can lead to chaos, as Rothman observed. We offer an alternative to the gauge formulation. This alternative is based on the fact that the vector potential is real and is defined by several physical requirements in addition to its curl. Our main results are listed below.

a. The relationship, $\nabla \cdot E_I = (\partial^2 \varphi_C / \partial t^2)/c^2$ is a previously unrecognized fundamental law of induction. It reflects the central role of retardation effects and supplements Faraday's law of induction, to permit a full definition of $E_I$, as well as $E_C$. It completes the set of Maxwell's equations and can properly be termed a missing Maxwell equation. The Lorenz equation is seen to be an alternative statement of this missing law.

b. Generally, two gauge choices must be considered rather than one. The hidden choice, $\varphi_I = 0$, must be recognized. Physically meaningless results occur when conflicting gauge choices are adopted.

c. As shown by Konopinski, the vector potential has both physical meaning and measurability as a field momentum. Feynman et al. discuss similar arguments, in the context of quantum mechanics, using the Bohm-Aharanov experimental results. The results of the present analysis are consistent with their conclusions that the potentials are physically real and central to the formalism.

d. The standard gauge is always employed in the presence of dynamic Coulomb fields. The gauge concept does not apply in the absence of dynamic Coulomb fields.

e. Imposing the requirement that the Coulomb field is only real field in the rest frame of a point charge leads to the selection of the standard gauge as the only physically meaningful choice.

## VII. Acknowledgements

I thank Mark A. Johnson for his technical assistance and for the support that made this work possible.

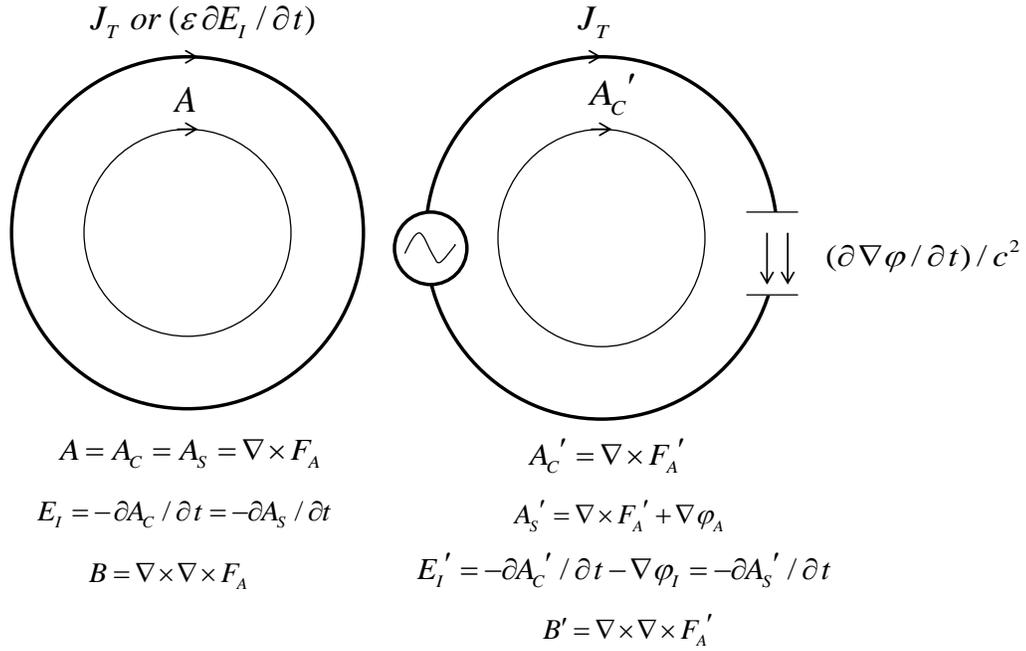

**Figure 1. Schematic of two general classes of problems**.

Figure 1a corresponds to closed current loops comprised of either true currents or displacement currents. Dynamic Coulomb fields do not exist so the gauge concept cannot apply.

Figure 1b corresponds to the case where dynamic Coulomb fields exist and gauge choices is required. The two gauge choices give $A_S'$ and $A_C'$, as illustrated.